\documentclass[%
reprint,
amsmath,amssymb,
prb,
]{revtex4-1}

\usepackage{graphicx}% Include figure files
\usepackage{dcolumn}% Align table columns on decimal point
\usepackage{bm}% bold math
\usepackage[bookmarks]{hyperref}% add hypertext capabilities
\usepackage[version=3]{mhchem}
\usepackage[all]{hypcap}
\usepackage{float}
\usepackage{amsmath}
\usepackage[usenames,dvipsnames,svgnames,table]{xcolor}
\usepackage[english]{babel}

\usepackage{xcolor}
\usepackage{soul}

\begin{document}
	
	%\preprint{APS/123-QED}
	
	\title{Charge-spin interconversion in graphene-based systems from density functional theory}
	
	\author{M. Rassekh$^{1,2}$}\email{maedeh.rassekh@estudiante.uam.es}
	\author{Hernán Santos$^{3}$}
	\author{A. Latg\'e$^{4}$}
	\author{Leonor Chico$^{5,6}$}
	\author{S. Farjami Shayesteh$^{1}$}
	\author{J. J. Palacios$^{2}$}\email{juanjose.palacios@uam.es}
	
	\affiliation{$^{1}$Department of Physics, University of Guilan, 41335-1914 Rasht, Iran\\
		$^{2}$Departamento de Física de la Materia Condensada, Condensed Matter Physics Center (IFIMAC), and Instituto Nicolás Cabrera (INC), Universidad Autónoma de Madrid, Cantoblanco 28049, Spain\\
		$^{3}$Dept. de Matem\'atica Aplicada, Ciencia e Ingenier\'{\i}a de Materiales y Tecnolog\'{\i}a Electr\'onica, ESCET, Universidad Rey Juan Carlos, C/ Tulip\'an s/n, M\'ostoles 28933, Madrid, Spain\\
		$^{4}$Instituto de F\' isica Universidade Federal Fluminense, Niter\' oi, Av. Litor\^ anea 24210-340, RJ-Brazil\\
		$^{5}$GISC, Departamento de Física de Materiales, Facultad de Ciencias Físicas, Universidad Complutense de Madrid, 28040 Madrid, Spain\\
		$^{6}$Instituto de Ciencia de Materiales de Madrid, Consejo Superior de Investigaciones Cient\'{\i}ficas, C/ Sor Juana In\'es de la Cruz 3, 28049 Madrid, Spain}

	\date{\today}% It is always \today, today,
	
	\begin{abstract}
		We present a methodology to address, from first principles, charge-spin interconversion in two-dimensional materials with spin-orbit coupling. Our study relies on an implementation of density functional theory based quantum transport formalism adapted to such purpose.  We show how an analysis of the $k$-resolved spin polarization gives the necessary insight to understand the different charge-spin interconversion mechanisms. We have tested it in the simplest scenario of isolated graphene in a perpendicular electric field where effective tight-binding models are available to compare with. Our results show that the flow of an unpolarized current across a single layer of graphene produces, as expected, a  spin separation perpendicular to the current for two of the three spin components (out-of-plane and longitudinal), which is the signature of the spin Hall effect. Additionally, it also yields an overall spin accumulation for the third spin component (perpendicular to the current), which is the signature of the Rashba-Edelstein effect. Even in this simple example, our results reveal an unexpected competition between the Rashba and the intrinsic spin-orbit coupling. Remarkably, the sign of the accumulated spin density does not depend on the electron or hole nature of the injected current for realistic values of the Rashba coupling. 
		
		%Results for actual systems, where the Rashba coupling is induced by proximity, which include graphene with Au adatoms and graphene in proximity to MoS$_2$, are also presented and discussed.  
	\end{abstract}
	
	\maketitle
	
	%\tableofcontents
	
	\section{\label{sec:level1} Introduction}
	
	Modern spintronics relies not only on the generation, manipulation and detection of spin-polarized currents through ferromagnetic
	materials but also on the increasingly important phenomena related to the interconversion of spin and charge currents. 
	This conversion can be made by an external electric field via the magnetoelectric coupling or, most importantly, via the spin-orbit coupling (SOC) \cite{sinova2015spin,manchon2015new,sander20172017,garello2018sot,garello2018sot}. The latter approach is typically associated with the Bychkov-Rashba or the Dresselhaus effects \cite{bychkov1984properties,GDressel1955}, which consist of a momentum-dependent splitting of spin bands. The Bychkov-Rashba splitting occurs in low-dimensional systems, such as semiconductor heterostructures or metal surfaces \cite{koroteev2004strong,ast2007giant,bihlmayer2007enhanced, mirhosseini2010toward}, whereas the Dresselhaus effect takes place in non-centrosymmetric bulk materials \cite{nikolaev2020skyrmionic}.
	
	In the paradigmatic spin Hall effect (SHE) \cite{valenzuela2006direct,sinova2015spin} a charge current flowing in a given direction generates a transverse spin current flowing in the perpendicular direction. The generation of this spin current leads to spin accumulations
	with opposite magnetizations at the edges of the sample. This phenomenon, which appears in electrically conductive and non-magnetic materials with a sizable SOC, can be experimentally measured through different techniques \cite{sinova2015spin,valenzuela2006direct}. In the Rashba-Edelstein effect (REE) \cite{edelstein1990spin,inoue2003diffuse} (or inverse spin galvanic effect), a non-equilibrium spin density is generated instead of spin current on account of a charge current. Likewise, a finite SOC is an essential ingredient. Both spin currents and densities can be used, for instance,  to manipulate the magnetization of ferromagnets.  Their reciprocal effects, the inverse SHE (ISHE) \cite{saitoh2006conversion} and inverse REE (IREE) \cite{sanchez2013spin,isasa2016origin, ganichev2002spin,Khokhriakov2020}), consist of the conversion of spin currents or spin densities, respectively, into charge currents.
	
	Many classes of materials are being explored in this context, including heavy element metals such as Au and Bi, oxides, and topological insulators (TIs) \cite{manchon2019current,PhysRevB.96.180415}. Of increasing relevance are two-dimensional (2D) crystals that essentially consist of atomically thin layers  obtained from van der Waals materials \cite{novoselov20162d} and whose
	electrical, optical, and spin properties can be easily modified by proximity. 
	Among these, graphene has played a prominent role in these investigations. Graphene presents the longest spin-relaxation length ever measured at room temperature \cite{ingla201524}, in part due to carbon being a light element with a relatively small intrinsic SOC \cite{min2006intrinsic}. However, this property precludes graphene from being able to play an active role in charge-spin interconversion. 
	In order to enhance its efficiency in this respect, there is an ongoing quest to increase the SOC splitting by external means, including mechanical deformation of the graphene lattice, applied electric fields, defects, addition of heavy adatoms that can increase the $d$-orbital contribution to SOC, as well as magnetic atoms that can induce an exchange-mediated spin splitting \cite{weeks2011engineering,eremeev2014spin,ma2012strong,marchenko2012giant,avsar2014spin,Santos2016,Santos2017}. 
	
	One of the most promising strategies consists of placing graphene in contact with other materials that can transfer their SOC by proximity. In fact, the observation of the SHE was originally reported in graphene decorated with adatoms and in contact with WS$_2$ \cite{avsar2014spin}. However, further studies have questioned 	the spin-related interpretation of the measured effects \cite{kaverzin2015electron,wang2015neutral,van2016spin}.
	More recently, the ISHE was observed in heterostructures comprising multilayer graphene and MoS$_2$ \cite{safeer2019room}. Even though the observation of the ISHE was unequivocal, it was not easy to discriminate	between proximity-induced SOC in graphene or in bulk MoS$_2$ because the latter 
	is believed to be conductive. This debate, however, seems to have been clarified in a recent work \cite{benitez2020tunable}. Moreover, recent experiments have demonstrated the spin galvanic effect (or IREE) in graphene/TI heterostructures \cite{Khokhriakov2020}.
	
	\begin{figure}[t]
		\includegraphics[width=1.0\columnwidth]{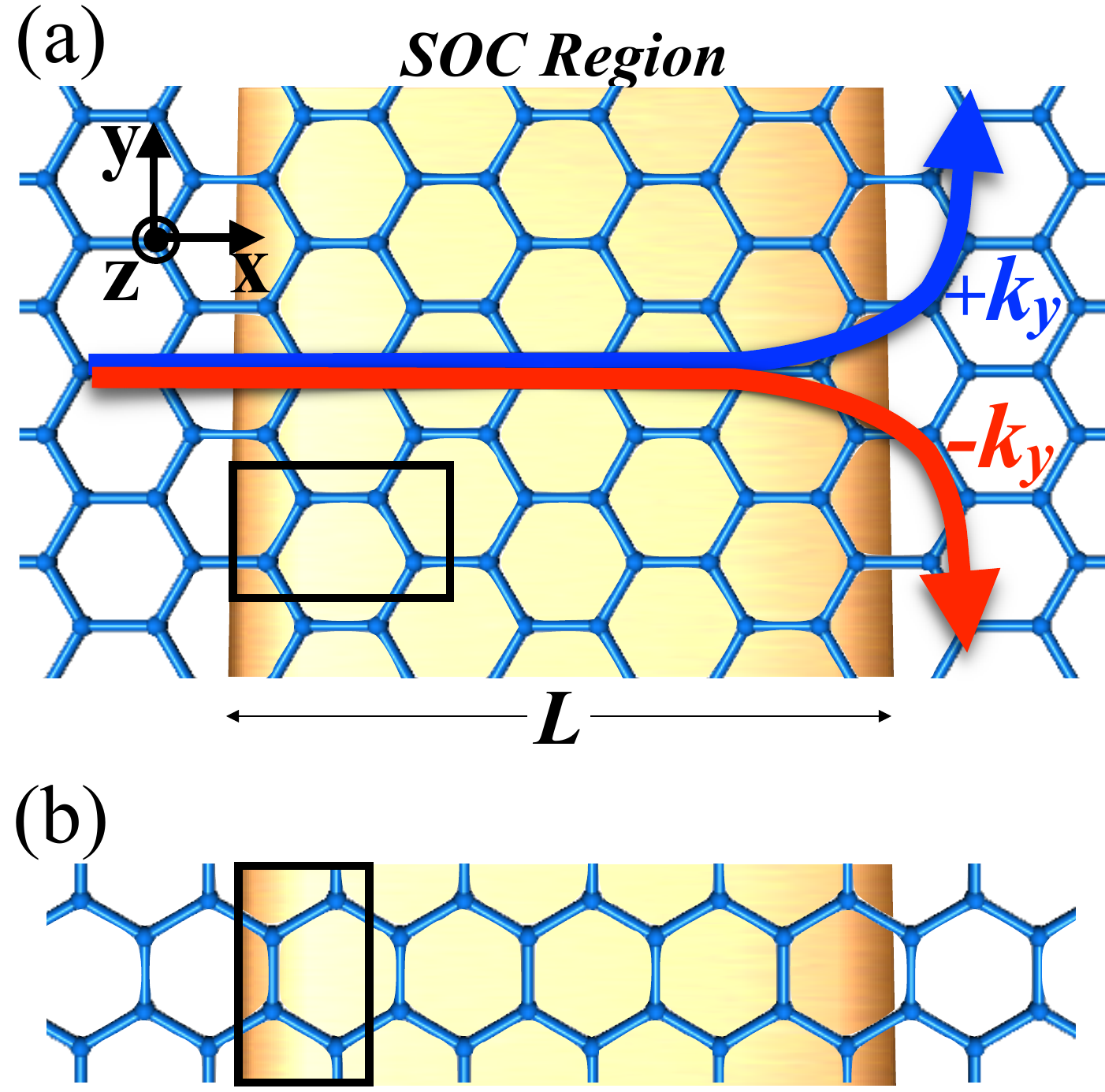}
		\caption{\small (Color online). Schematic view of our proposed graphene setup for zigzag (a), and armchair (b) orientations.  An infinite graphene flake is divided into three regions, the central one (of length $L$ and shown in yellow) being subject to a strong electric field to induce a Rashba effect therein. Black boxes show the unit cells used as the units of this length in the text. The lateral areas are SOC-free zones. The red and blue arrows represent how opposite spin carriers (along the $x$-quantization axis) deviate in opposite directions resulting in the SHE. }
		\label{schematics}
	\end{figure}
	
	On the theoretical side, density functional theory (DFT) studies have supported the possibility that  proximity-induced SOC can enhance the graphene charge-spin interconversion  efficiency \cite{gmitra2016trivial,wang2015strong,gmitra2015graphene,offidani2017optimal,milletari2017covariant}.  The information obtained from the DFT calculations can be transferred  to effective models, which can then be used to compute charge-spin related quantities. These calculations are typically based on Kubo formalism \cite{garcia2017spin} or non-equilibrium Green's function formalism \cite{nikolic2005mesoscopic,nikolic2005nonequilibrium}, all of them being real-space methodologies. Since the subtle proximity effects induced, e.g., by transition metal dichalcogenides (TMD) on graphene can only be captured by DFT calculations, it would be desirable to make a direct connection between these calculations and charge-spin interconversion-related quantities without the need of an effective model, not always at hand. 
	
	In this work we present a DFT-based quantum transport methodology to quantitatively address charge-spin interconversion in 2D crystals. The novelty here relies on the $k$-dependent evaluation of the spin polarization of the current, where $k$ is the crystal momentum perpendicular to the direction of the current. This allows us to obtain the charge-spin fingerprint in reciprocal space for any material or combination of materials directly from DFT. We have tested our methodology on free-standing graphene with Rashba coupling induced by the presence of a strong perpendicular electric field. For this paradigmatic system a simple tight-binding (TB) model is available which we use to verify the results obtained from our DFT implementation and to aid in the interpretation of these. 
	
	%We have also addressed realistic systems such as graphene with Au adatoms and the hybrid system MoS$_2$/graphene for which several experiments have shown its potential.
	
	%Graphene with its unique properties such as high two-dimensional carrier’s mobility \cite{morozov2005two} has proven to be a truly wonder material over the past decade. It is considered as a promising element in next-generation electronic and spintronic devices, e.g., the quantum Hall effect \cite{zhang2005experimental} and the spin Hall effect \cite{liu2007intrinsic}. In consequence, it is investigated both in the experiment and in theory with great effort.
	
	\section{\label{sec:level2} Theoretical approach}
	
	\subsection{Methodology: k-dependent spin polarization}
	
	A schematic view of the proposed graphene-based device is shown in Fig. \ref{schematics}. The system consists of a single and infinite layer of graphene with a central region where the Rashba coupling is induced either by a perpendicular electric field, the presence of heavy adatoms or by proximity to other materials. This way, the graphene flake is divided into three regions: the central SOC active region,  where the charge-spin interconversion takes place and that here is infinite in the $y$ direction, and the left and right semi-infinite regions, which inject and collect the current, respectively. Since the system is infinite in the direction perpendicular to the current, which flows in the
	$x$ direction, the wavevector $k_y$ (parallel to the interface) is a good quantum number. 
	This magnitude, combined with the knowledge of the band structure, can be used to single out the left-going and right-going contributions to the total current. 
	
	\begin{figure}[ht!]
		\includegraphics[width=1.0\columnwidth]{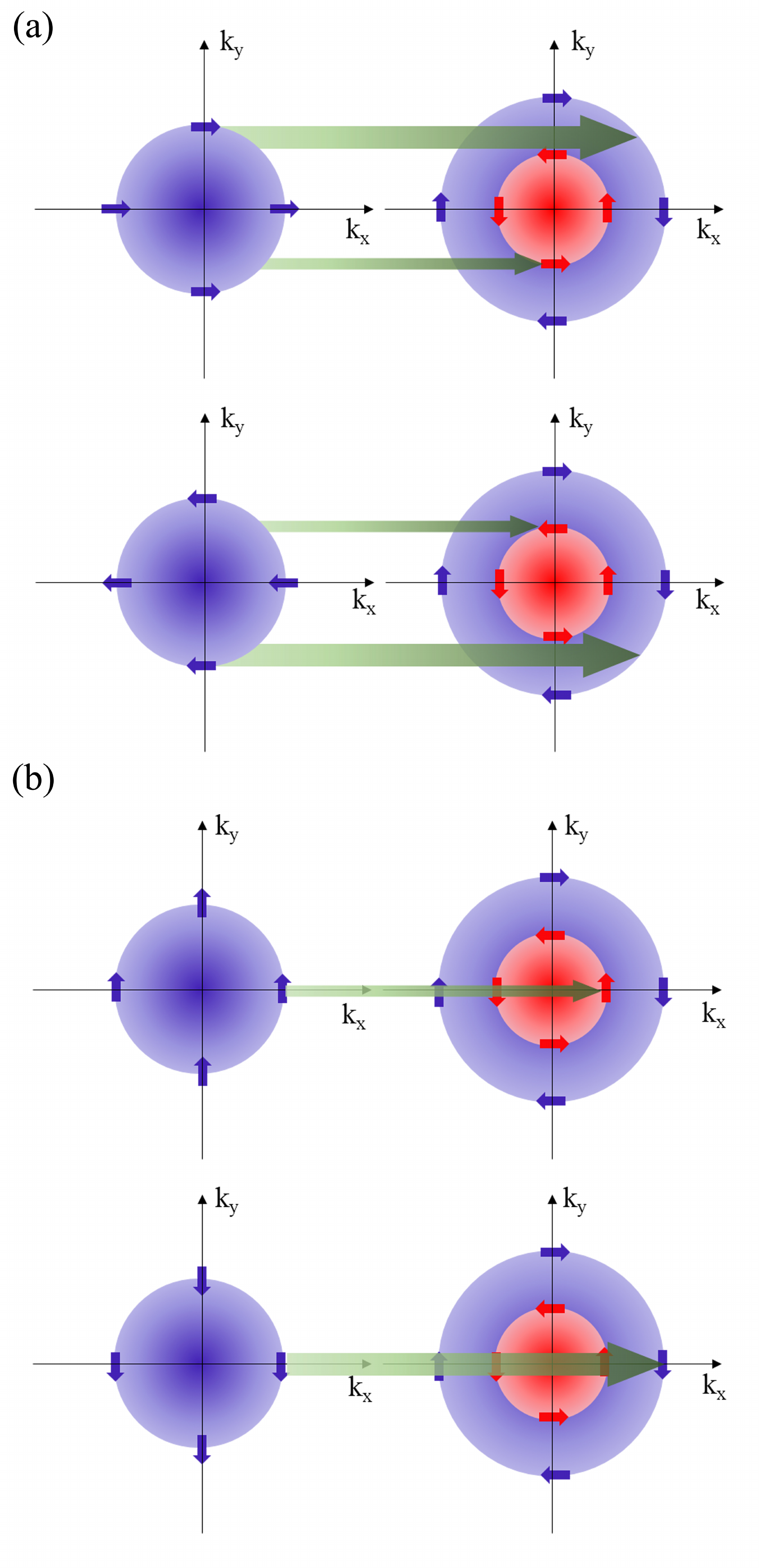}
		\caption{\small (Color online).Schematic representation of the origin of the SHE (a) and REE (b) in graphene for transport along the positive $x$ axis. The Fermi level is set to positive energies (electrons) and, therefore, only states with positive velocities on the right semicircles are considered. The blue arrows indicate the spin of the injected carriers.}
		\label{fig:simplepicture}
	\end{figure}
	
	Charge-spin interconversion mechanisms (SHE and REE) can be easily identified and quantified by the spin polarization of the $k_y$-dependent current, as discussed below; notwithstanding, a simple picture for the origin of these effects can already be anticipated from the schematics in Fig. \ref{fig:simplepicture}. There we show two situations giving rise to the SHE (a), and the REE (b), respectively, for the simplest Rashba coupling scenario. The Fermi surface of the graphene region injecting the current is depicted on the left of the figures, whereas the spin-split Fermi surface of the graphene region with Rashba coupling is shown on the right (we only present one valley, since both valleys behave identically with respect to the spin splitting of the bands \cite{Rashba.PhysRevB.79.161409}). The green arrows connect the sections of the Fermi surfaces between which the current is expected to be significant for different spin orientations of the incoming electrons. From panel (a) (SHE) one anticipates a predominant current for positive (negative) $k_y$ and parallel (antiparallel) orientation of the injected spins along $x$ \cite{Dyrdal.PhysRevB.80.155444}. In panel (b) (REE), there is no left-right asymmetry in the injected current, but this is expected to be polarized due to the different transmission of opposite-spin injected carriers \cite{Dyrdal.PhysRevB.89.075422}. When the spin polarization of the injected current is perpendicular to the plane, a SHE may also emerge, but such a simple picture is no longer available.
	
	To properly quantify both effects we define the spin polarization of the current at a given energy $E$ as in Ref. \onlinecite{Chico2015}, 
	\begin{equation}
	\label{polarization}
	P=T_{\upuparrows}+T_{\downarrow\uparrow}-T_{\uparrow\downarrow}-T_{\downdownarrows}\,\,,
	\end{equation}
	where the spin-conserved ($T_{\upuparrows}$, $T_{\downdownarrows}$) and the spin-flip ($T_{\uparrow\downarrow}$, $T_{\downarrow\uparrow}$) transmission functions are obtained with the help of the Green's function of the SOC-active region and the coupling matrices to left (L) and right (R) regions \cite{jacob2011critical,Chico2015}
	\begin{equation}
	\label{caroli}
	T_{\sigma \sigma^{\prime}}=\operatorname{Tr}\left[\boldsymbol\Gamma_{\sigma}^{L} \boldsymbol{G}_{\sigma \sigma^{\prime}} \boldsymbol\Gamma_{\sigma^{\prime}}^{R} \boldsymbol{G}_{\sigma \sigma^{\prime}}^{\dagger}\right].
	\end{equation}
	Therefore, more specifically, we must compute
	\begin{equation}
	\begin{split}
	P
	=&Tr\left[\boldsymbol\Gamma_{\uparrow}^L \boldsymbol{G}_{\upuparrows} \boldsymbol\Gamma_{\uparrow}^R \boldsymbol{G}_{\upuparrows}^\dagger+ 
	\boldsymbol\Gamma_{\downarrow}^L \boldsymbol{G}_{\downarrow\uparrow} \boldsymbol\Gamma_{\uparrow}^R \boldsymbol{G}_{\downarrow\uparrow}^\dagger -
	\right.
	\\
	&\left.-\boldsymbol\Gamma_{\uparrow}^L \boldsymbol{G}_{\uparrow\downarrow} \boldsymbol\Gamma_{\downarrow}^R \boldsymbol{G}_{\uparrow\downarrow}^\dagger - \boldsymbol\Gamma_{\downdownarrows}^L \boldsymbol{G}_{\uparrow\downarrow} \boldsymbol\Gamma_{\downarrow}^R \boldsymbol{G}_{\downdownarrows}^\dagger\right].
	\end{split}
	\end{equation}
	As such, these equations imply that spin is conserved in the electrodes, with $\uparrow$ and $\downarrow$ denoting spin eigenstates along an arbitrary direction (here $x,y,$ and $z$). We will refrain from integrating in energy since a small bias voltage is assumed and we are only interested in linear effects.
	
	For the DFT implementation of the quantum transport calculation presented in this work, the Green's function (with spin indices omitted here) is defined as
	
	\begin{equation}	
	\bm{G}(E)=[z\bm{S}-\bm{H}_{\rm C}-\bm{\Sigma}_T(E)]^{-1}\,\,,
	\label{Greens}
	\end{equation}
	with $z=E+ i\eta $ and $\eta$ being an infinitesimal number.  $ \bm{H}_{\rm C} $ is the Hamiltonian of the central part where SOC is present,  $ \bm{S} $ the corresponding overlap matrix for the basis functions, which are assumed not necessarily orthogonal, and $ \bm{\Sigma}_T (E) $ is the summation of right and left self-energies from which the coupling matrices in Eq. \ref{caroli} are obtained:
	\begin{equation}
	\begin{split}
	& \bm\Gamma^L (E)= i \left( \bm{\Sigma}_L (E) - (\bm{\Sigma}_L (E))^\dagger\right) ,
	\\
	& \bm\Gamma^R (E)= i \left( \bm{\Sigma}_R (E) - (\bm{\Sigma}_R (E))^\dagger\right) .
	\end{split}
	\end{equation}
	The right and left self-energies, which are identical here,  are calculated from the on-cell Hamiltonian, $\bm H_0$, and corresponding overlap matrix, $\bm S_0$,  by the following Dyson equations:
	\begin{eqnarray}
	\nonumber
	\begin{split}
	& \bm{\Sigma}_L (E) = (\bm{H}_1^\dagger-z\bm{S}_1^\dagger) (z\bm{S}_0-\bm{H}_0-\bm{\Sigma}_L(E))^{-1} (\bm{H}_1-z\bm{S}_1)
	\\
	& \bm{\Sigma}_R (E) = (\bm{H}_1-z\bm{S}_1) (z\bm{S}_0-\bm{H}_0-\bm{\Sigma}_R(E))^{-1} (\bm{H}_1^\dagger-z\bm{S}_1^\dagger),
	\end{split}
	\end{eqnarray}
	where $\boldsymbol{H}_1$ and $\boldsymbol{S}_1$ are the forward hopping and overlap matrices, respectively. 
	
	In the transmission calculations the area where SOC is active can cover all the central region defined through $H_C$ in Eq. \ref{Greens} or just part of it to allow for SOC-free buffer zones and a smoother or perfect electronic match with the SOC-free self-energies (see Fig. \ref{schematics}). Although we have not appreciated differences between these two possibilities in the results, including buffer zones is safer and our DFT spin polarization results are obtain in this way. Also, in order to conserve spin in the electrodes, we remove SOC from them, but we keep the effect of the electric field. When SOC is absent from $H_{\rm C}$, the transmission is perfect and no spin polarization or interference effects appear.

\begin{figure}[b]
		\includegraphics[width=1.0\columnwidth]{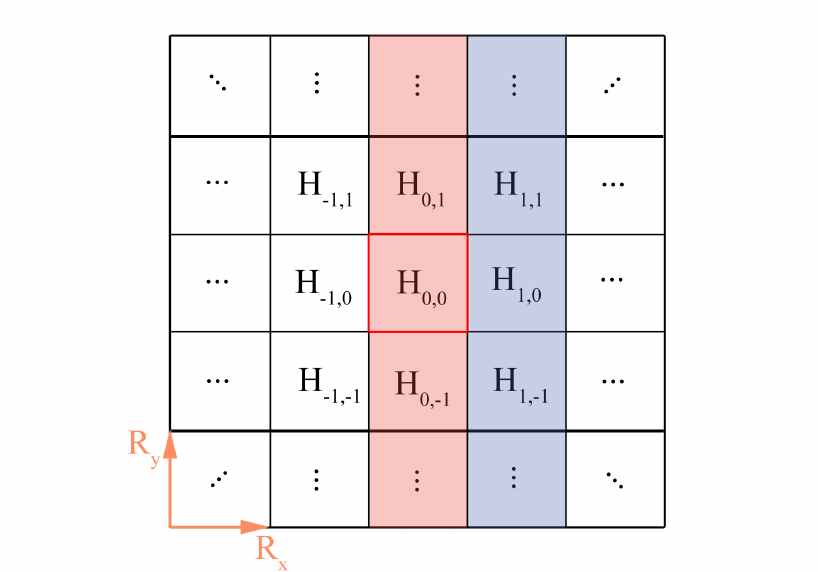}
		\caption{\small Schematic illustration of Hamiltonian/overlap matrix partitioning.}
		\label{Hamiltonian}
	\end{figure}

	The above-presented methodology is standard in quantum transport, but the infinite nature of our system along the parallel ($y$) direction allows to replace the Hamiltonian and overlap matrices with their lattice Fourier transform only in the $y$ direction, so $ \bm{H}_0$ and $ \bm{H}_1 $ are actually given by (see Fig. \ref{Hamiltonian})
	
	\begin{figure*}[t!]
		\centering
		\includegraphics{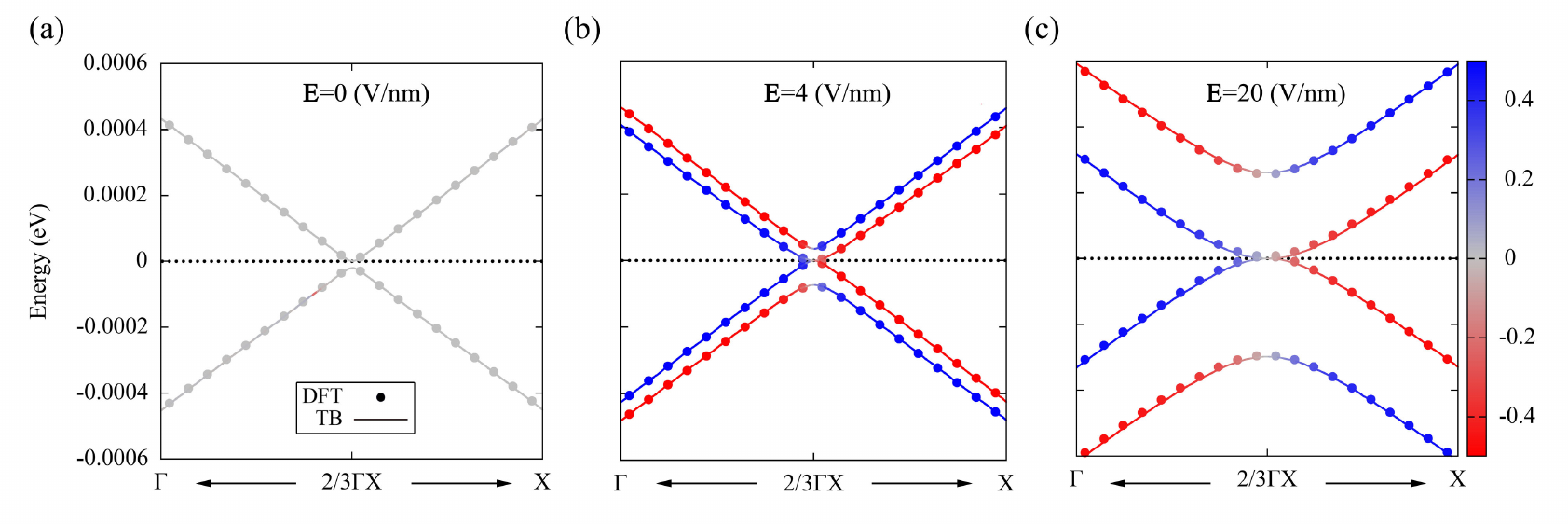}
		\caption{\small DFT and TB bands near one of the Dirac points, without electric field (a), with an external electric field of 4 V/nm (b), and 20 V/nm (c). The color bar indicates the $y$-component of the spin expectation value.}
		\label{Bands}
	\end{figure*}

	\begin{equation}
	\begin{split}
	& \bm{H}_0=...+\bm{H}_{0,-1}e^{-ik_yR_y}+\bm{H}_{0,0}+\bm{H}_{0,1}e^{ik_yR_y}+...
	\\
	& \bm{H}_1=...+\bm{H}_{1,-1}e^{-ik_yR_y}+\bm{H}_{1,0}+\bm{H}_{1,1}e^{ik_yR_y}+...
	\end{split}
	\end{equation}
	or in general
	\begin{equation}
	\bm{H}(R_x,k_y)=\sum_{R_y}H(R_x,R_y)e^{ik_yR_y}\,\,,
	\end{equation}
	where $\boldsymbol{R}$ are Bravais lattice vectors. Similarly,
	\begin{equation}
	\bm{S}(R_x,k_y)=\sum_{R_y}S(R_x,R_y)e^{ik_yR_y}\,\,.
	\end{equation}

	These equations finally result in a $k_y$-dependent spin polarization function $P(E,k_y)$ that contains all the needed information. We note that a related methodology was proposed in Ref. \onlinecite{Dolui.PhysRevB.96.220403}, although in a different context where the focus was on the spectral function, not the spin polarization.

	\subsection{Atomistic Modelling}
	The matrices needed for the evaluation of Green's functions, as explained in the previous section, are obtained from the DFT implementation of the OpenMX code \cite{OpenMX,ozaki2003variationally,ozaki2004numerical,ozaki2005efficient,lejaeghere2016reproducibility,lee2019openmx}. This code is based on the norm-conserving pseudo-potential method \cite{bachelet1982pseudopotentials,troullier1993pseudopotential,vanderbilt1990soft,blochl1990generalized,morrison1993nonlocal} with a partial core correction and pseudo atomic orbitals (LCPAO) as basis functions \cite{ozaki2003variationally,ozaki2004numerical} here specified by $ C(6.0)-s^2p^2d^1 $. Where $ C $ is the atomic symbol for Carbon, 6.0 is the cutoff radius in units of Bohr, and $ s^2p^2d^1 $ indicates that two primitive orbitals for each of $ s $ and $ p $ components and one primitive orbital for $ d $ components are employed. The fully relativistic effects, including spin-orbit coupling (SOC), have been included in the non-collinear DFT calculations via the j-dependent pseudo-potentials scheme \cite{macdonald1979relativistic,bachelet1982pseudopotentials,theurich2001self}.
	The exchange-correlation functional chosen here is the spin-polarized GGA-PBE \cite{perdew1996generalized}. All calculations were performed until the change in total energy between two successive iteration steps converged to less than 10$^{-6}$ Hartree. A cutoff energy of 220 Ry and a $ 7\times7\times1 $ $k$-grid have been used in all results presented below. 
	%Van der Waals interaction (vdW) correlation is considered by using the semiempirical dispersion-corrected density functional theory (DFT-D2) force-field approach. 
	A vacuum spacing of 20 \AA\ in the $z$ direction is used to prevent the interaction between periodic images. Atomic SOC in the presence of a perpendicular electric field, adatoms, or other materials in proximity, gives rise to both intrinsic and Rashba couplings.In our implementation we remove the SOC terms from the lateral buffer zones and obtain the matrices to compute the self-energies from there.  Since the effect of the electric field  remains, the electronic mismatch with the SOC-active region is reduced to a minimum.
	
	To test our DFT implementation and elucidate the role of the two SOC contributions, we have also considered a simple TB model for graphene-based on four atomic orbitals ($s, p_x, p_y, p_z$). SOC effects are included by adding the atomic term $H_{\text {SOC }}$ to the TB Hamiltonian \cite{Santos2013}. With the usual assumption that the most important contribution of the crystal potential to the SOC is close to the cores,  $H_{\text {SOC }}$ is given by 
	$$
	H_{\mathrm{SOC}}=\sum_{i} \frac{\hbar}{4 m^{2} c^{2}} \frac{1}{r_{i}} \frac{d V_{i}}{d r_{i}} \boldsymbol{L} \cdot \boldsymbol{S}=\lambda_I \boldsymbol{L} \cdot \boldsymbol{S},
	$$
	where  $V_{i}$ is assumed to be spherically symmetric; $r_{i}$ is the radial coordinate with origin at the $i$-th atom;  $\boldsymbol{L}$ is the electron orbital angular momentum operator, and $\boldsymbol{S}$ is the spin operator. The parameter $\lambda_I$ is a renormalized atomic SOC constant that depends on the angular momentum. Note that $H_{\text {SOC }}$ only couples $p$ orbitals in the same atom. Since spin is included, the Hamiltonian matrix has $8 N_{a} \times 8 N_{a}$ elements, where $N_{a}$ is the number of carbon atoms in the unit cell, and 8 corresponds to the four orbitals per spin of the basis set. Therefore, the total Hamiltonian in the $2 \times 2$ block spinor structure is 
	$$
	H=\left(\begin{array}{cc}
	H_{0}+\lambda_I L_{z} & \lambda_I\left(L_{x}-i L_{y}\right) \\
	\lambda_I\left(L_{x}+i L_{y}\right) & H_{0}-\lambda_I L_{z}
	\end{array}\right).
	$$
	
	As in the DFT description, the presence of a perpendicular electric field can be directly added to the four-orbital Hamiltonian. However, we choose to take its effect into account through the effective one-orbital Hamiltonian \cite{Zhang2014,Chico2015,Santos2020} 
	\begin{equation} 
	H_{\rm R}= %\gamma_0 \sum_{\substack{<i,j>\\\sigma}} c_{i\alpha}^\dagger  c_{j\alpha} + 
	\frac{i \lambda_R}{a_{cc}}\sum_{\substack {<i,j>\\\sigma,\sigma^\prime}} c_{i\sigma}^\dagger  
	\big[ (\boldsymbol{\sigma} \times \bold{d}_{ij}) \cdot \bold {e}_p \big]_{\sigma \sigma^\prime} 
	\ c_{j\sigma^\prime} \,\,,
	\label{HR} 
	\end{equation}
	acting only on $p_z$ orbitals with $\boldsymbol{\sigma}$ being the Pauli spin matrices in vector notation, $\bold{d}_{ij}$ the position vectors between atoms $i$ and $j$,  $\lambda_R$ the Rashba strength, and $\bold {e}_p$ the unitary vector perpendicular to the plane of graphene.  The nearest-neighbor carbon-carbon distance in graphene is $a_{cc}=1.42$ \AA. The values  of the hopping parameters in our TB model, $\lambda_I$, and $\lambda_R$ are chosen to fit the DFT bands. The fitted values are $ \lambda_I=9.2 $ $ \mu$eV,	and $ \lambda_R= 18.1 $ and $ 93.3$ $\mu$eV for 4 and 20 V/nm electric fields, respectively. These values are in good agreement with those reported in Refs. \cite{Gmitra.PhysRevB.80.235431,abdelouahed2010spin}. Notice that, strictly speaking, the Rashba parameter $\lambda_R$ depends on $\lambda_I$ since we have included the full atomic SOC, $H_{\rm SOC}$, in our TB Hamiltonian. However, in our single-orbital model \ref{HR}, $\lambda_R$ can be independently fitted. Alternatively, one could have used a one-orbital effective Hamiltonian all along as in Ref. \cite{Gmitra.PhysRevB.80.235431}. As illustrated in Fig. \ref{Bands}, where the bands and their $y$-component of the spin expectation value (encoded in the color of the lines and dots) are shown near one of the Dirac points, the coincidence between DFT and TB results is	excellent up to the highest values considered for the electric field. The gaps also follow the evolution with the field previously reported \cite{Gmitra.PhysRevB.80.235431}.
	
	\begin{figure}[hb!]
		\includegraphics[width=0.98\columnwidth]{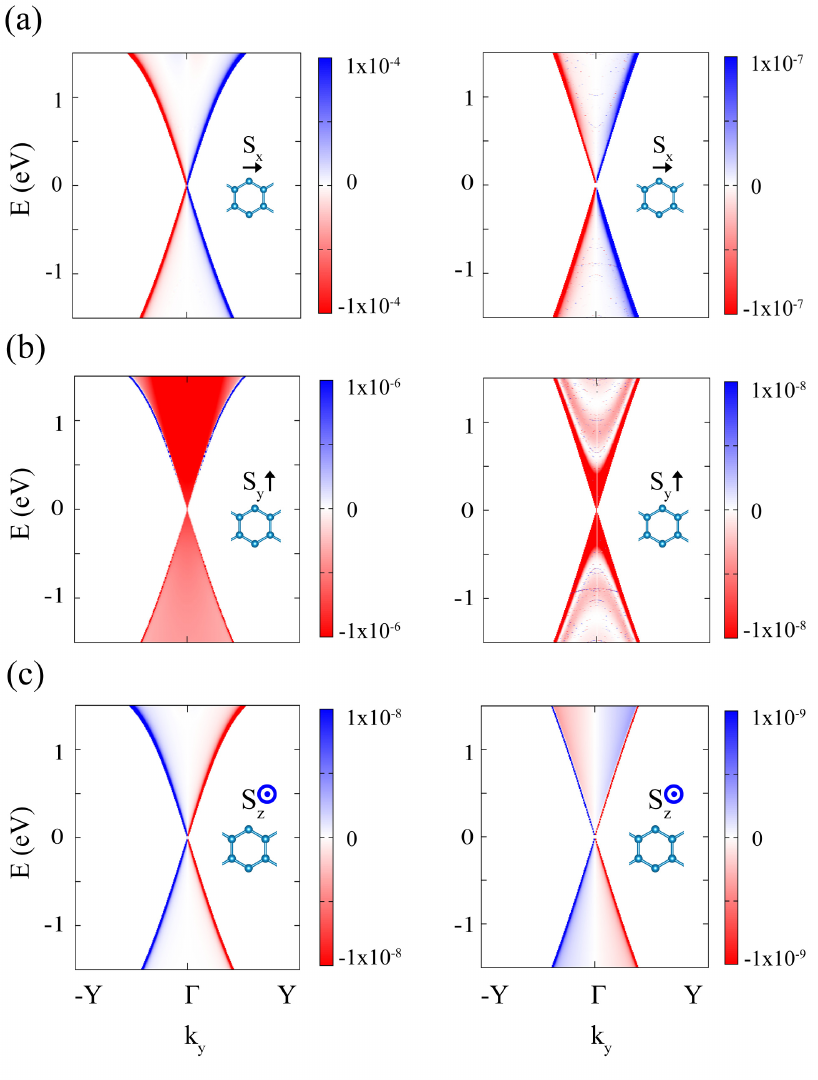}
		\caption{\small (Color online) Polarization for the armchair orientation. DFT and TB results are shown on the left and right columns, respectively. Results on (a), (b) and (c) correspond to the polarization for injected electrons with spin along the $x$, $y$, and $z$ directions, respectively. The length of the Rashba region is L-12 (L-n being the number of n unit cells, where the length of one unit cell (L-1) is 2.46 \AA) and the value of the electric field is 4 V/nm.}
		\label{polarization:armchair}
	\end{figure}	
	
	\begin{figure}[t!]
		\centering
		\includegraphics[width=0.98\columnwidth]{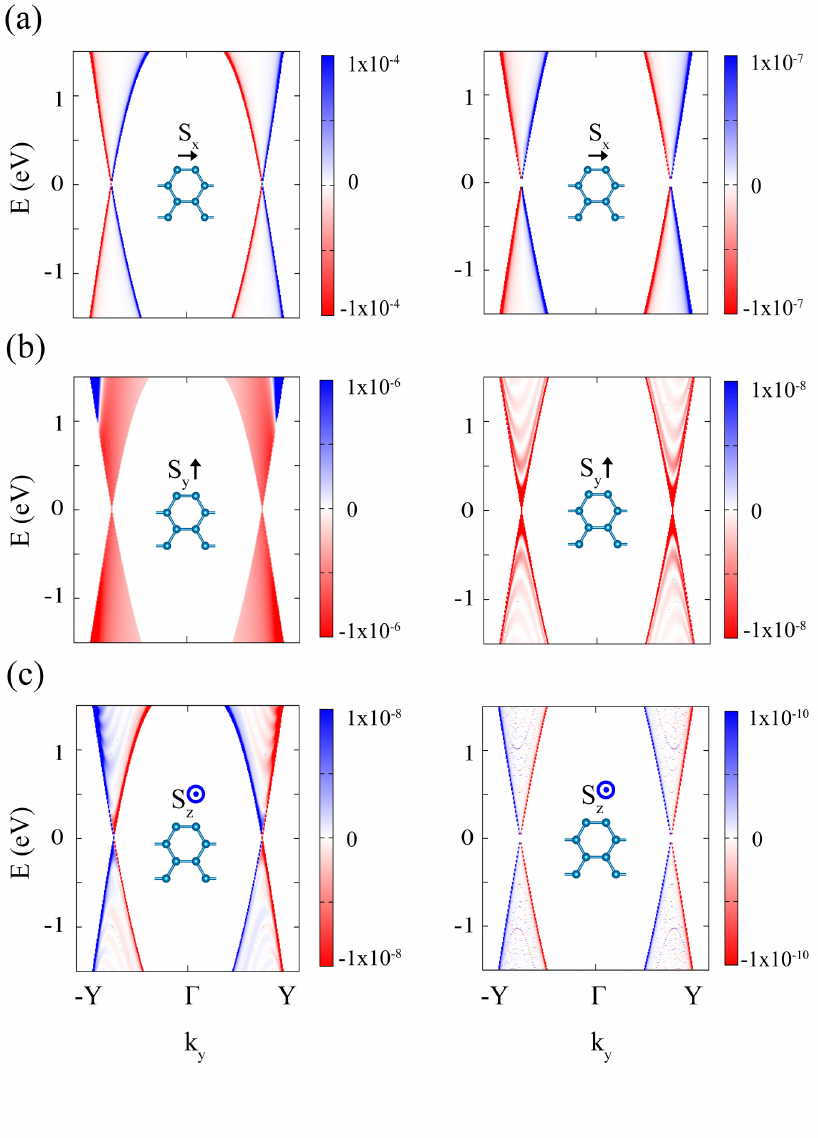}
		\caption{\small (Color online)   Polarization  for the zigzag orientation. DFT and TB results are shown on the left and right columns, respectively. Results on (a), (b) and (c) correspond to the polarization for injected electrons with spin along the $x$, $y$, and $z$ directions, respectively. The length of the Rashba region is L-12 (L-n being the number of n unit cells, where the length of one unit cell (L-1) is 4.26 \AA) and the value of the electric field is 4 V/nm.}
		\label{polarization:zigzag}
	\end{figure}
	
	\begin{figure*}[ht!]
		\centering
		\includegraphics{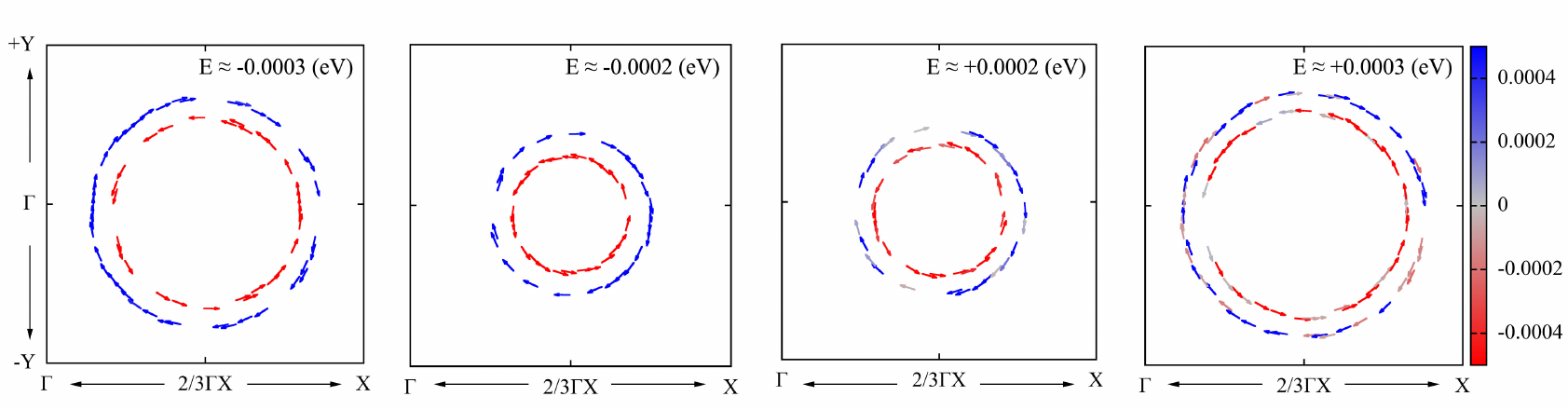}
		\caption{\small (Color online) DFT Fermi surface and spin texture for graphene in a perpendicular electric field of \textbf{E}=4 V/nm. Four different values of the Fermi energy near the Dirac point have been selected. The color bar refers to the $z$-spin projection which is too small to visibly reflect in the length of the arrow.}
		\label{BandSpin}
	\end{figure*}
	
	\begin{figure}[b!]
		\centering
		\includegraphics[width=1\columnwidth]{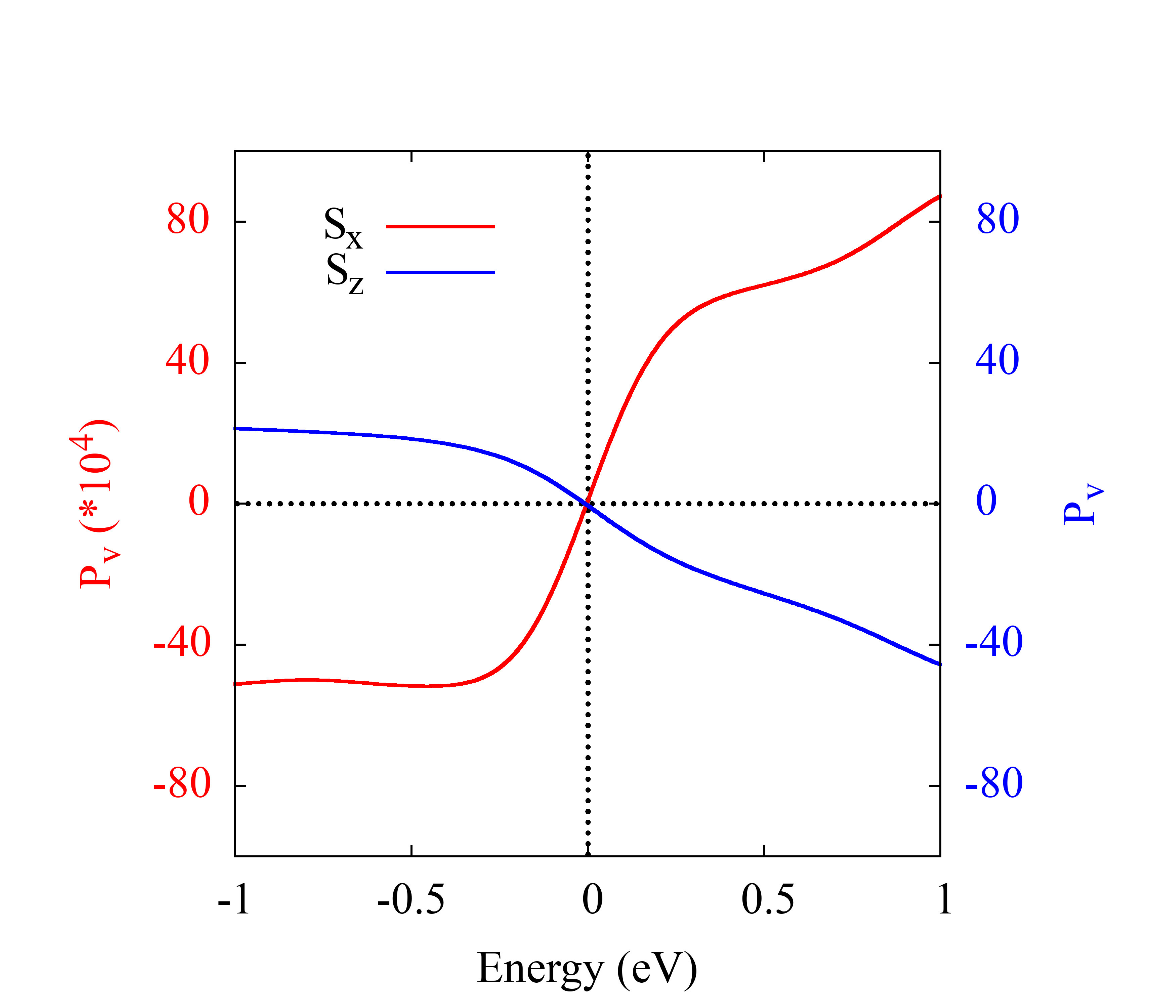}
		\caption{\small (Color online) Integrated spin polarization values over $ k_y $, as a function of energy for armchair orientation of DFT for the injected electrons with spin along the x, and z directions.}
		\label{pv}
	\end{figure}
	
	\section{Results and discussion}
	%\subsection{Graphene in a perpendicular electric field}
	
	Polarization results for graphene in a perpendicular electric field, obtained from the TB model and from DFT, are shown in Fig. \ref{polarization:armchair} and Fig. \ref{polarization:zigzag}.  We consider two orientations of the graphene plane, that we label armchair and zigzag, respectively; the orientation refers to the direction of the interface, which is perpendicular to that of the current, injected here in the $x$-direction, as mentioned before (see insets of Figs. \ref{polarization:armchair} and  \ref{polarization:zigzag}). For instance, the setup shown in Fig. \ref{schematics}(a) corresponds to the zigzag orientation.	A large value of the electric field of 4 V/nm has been chosen, which gives rise to a significant splitting of the bands, as shown in the middle panel in Fig. \ref{Bands}. It is important to mention that electric field values of this order correspond, in fact, to realistic situations. We can mention that a large interfacial electric field equal to 5 V/nm (or 0.5 V/{\AA}) have been reported on liquid/solid interfaces with high-density carrier accumulation \cite{yuan2013zeeman}. Also, an out-of-plane electric field of 1 V/{\AA} = 10 V/nm was addressed by {\it Houssa et al.} \cite{houssa2015silicene} and there are many references on electric fields of such magnitude applied on 2D materials \cite{gmitra2016trivial,rong2015chemical}. The three orthogonal spin components have been explored as indicated in the corresponding panels of Figs. \ref{polarization:armchair} and \ref{polarization:zigzag}. The energy origin has been set to the Dirac point in both, TB and DFT cases. The wave vector $k_y$ runs over the whole Brillouin zone.
	
	Blue and red colors represent positive and negative values of the spin polarization, respectively. From the sign of the spin polarization in Fig. \ref{polarization:armchair}(a), we conclude that spin-up and spin-down electrons (in the $x$-direction) are preferentially deflected in opposite spatial directions since the spin polarization sign follows that of the transverse group velocity (for this orientation the two Dirac cones are superimposed and behave identically in this regard). This result was already anticipated and discussed with the help  of Fig. \ref{fig:simplepicture}(a) and is the basis of the SHE. The same number of spin-up and spin-down electrons are moving in opposite directions  while the total transmission, 	$T=T_{\upuparrows}+T_{\downarrow\uparrow}+T_{\uparrow\downarrow}+T_{\downdownarrows}$, not shown here, is $k_y$-symmetric. Therefore, the net transverse charge current is 0, while a finite spin $S_x$ current flows in the transverse direction. In an actual sample with boundaries, this spin current would result in a spin accumulation of opposite signs on the lateral edges. Notably, the TB and DFT results are qualitatively coincident, which gives us confidence in our DFT-based implementation. In all cases, the spin polarization is maximum for the largest values of $|k_y|$  and zero for $k_y=0$ (as expected from the schematics in Fig. \ref{fig:simplepicture}), but shows a sub-structure that originates from interference due to the finite length of the SOC-active area (more clearly seen in the TB results). Also, not surprisingly, similar results are obtained when choosing the zigzag orientation, as shown in Fig. \ref{polarization:zigzag}. Now the two Dirac cones can be differentiated, becoming clear that they contribute in the same manner to the spin polarization.
	
	When the spin projection of the injected current is perpendicular to the graphene plane ($z$-direction), a similar $k_y$-dependent spin polarization can be seen in our results, albeit with reversed sign [see Figs. \ref{polarization:armchair}(c) and \ref{polarization:zigzag}(c)], being visibly smaller in absolute value than that of the $x$-projection, but not zero. The small spin polarization found in this case is understandable, since the $S_z$ component of the spin is only (and barely) appreciable near the Dirac point, as shown in the panels of Fig. \ref{BandSpin} for quite small energy values.
	This is in contrast with the results in more experimentally relevant situations \cite{safeer2019room},
	but is in agreement with previous theoretical works of Nikolić {\it et al.} on a two-dimensional electron gas  \cite{nikolic2005mesoscopic,nikolic2005nonequilibrium}. Our methodology does not allow us to make a direct comparison with experimental observations or with the spin conductance as defined, e.g., by Nikolić {\it et al.}  \cite{nikolic2005mesoscopic,nikolic2005nonequilibrium},  but we can get some intuition defining the quantity
		$$ P_v= \sum_{k_y}P(k_y,E)v_y(k_y,E), $$		
		which is shown in Fig. \ref{pv}. This quantity reflects now  the overall lateral separation of the spin polarization as a function of energy in close analogy with the spin current. Notice the lack of electron-hole symmetry expected in an simple TB model, but absent in the DFT calculations.
	
	As far as the $y$ spin component of the injected current is concerned, this shows spin accumulation [Figs. \ref{polarization:armchair}(b) and \ref{polarization:zigzag}(b)] since the spin polarization is symmetric with respect to $k_y$ (as is the total transmission, not shown). This result is at the heart of the REE. Naively, according to our simple picture discussed above (see Fig. \ref{fig:simplepicture}), the spin accumulation should change sign as the energy of the injected electrons crosses the Fermi level. However, this is not the case in either type of calculation, DFT or TB. (The fact that both DFT and TB results are essentially similar rules out any implementation problem or lack of accuracy of the DFT calculations as the origin of this unexpected result.) We attribute this result to the competing effect of the intrinsic and extrinsic (Rashba) couplings, which are still comparable in magnitude for the value of the  electric field we have considered \cite{Dyrdal.PhysRevB.80.155444,Correa2020}. Our DFT results in Figs. \ref{polarization:armchair} and \ref{polarization:zigzag} show that the sign of the spin polarization actually changes for electrons (positive energies), but away from the Fermi energy and only for states with wave vectors almost parallel to the interface. This change becomes more clear for larger values of the electric field such as 20 V/nm  and larger SOC-active areas (results not shown here).
	
\begin{figure}[ht!]
		\centering
		\includegraphics[width=0.98\columnwidth]{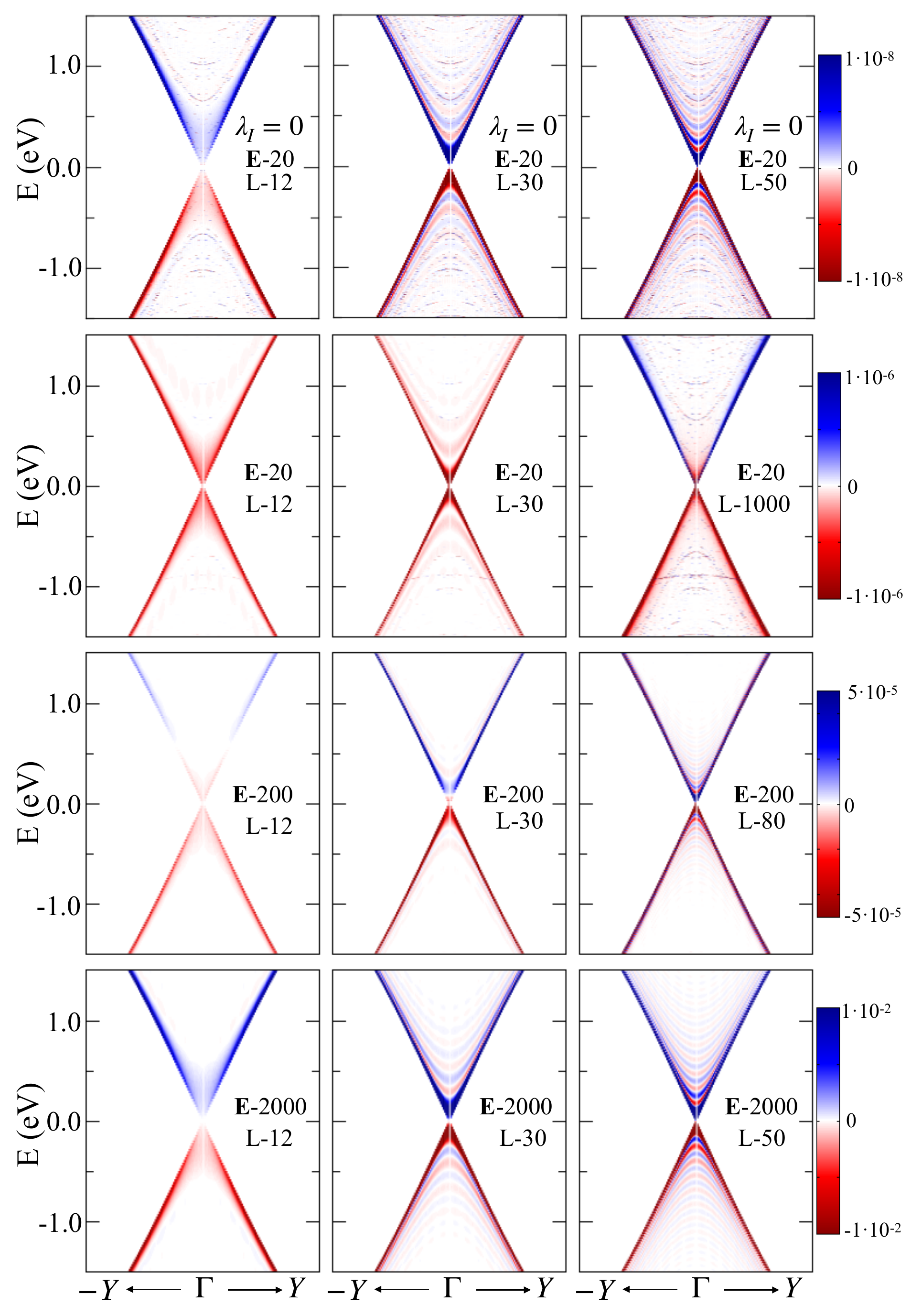}
		\caption{\small (Color online) Polarization results for the armchair orientation and injected electrons with spin along the $y$-direction following the tight-binding approach. Increasing electric field values are considered from top to bottom panels, from E = 20 to 2000 V/nm (\textbf{E}-20, \textbf{E}-200, and \textbf{E}-2000). Wider SOC areas are considered from left to right panels, L-n being the number of n unit cells, where the length of one unit cell (L-1) is 2.46 \AA. First row is calculated with $\lambda_I=0$, the others with the two SOC contributions, $\lambda_I$ and $\lambda_R$.}
		\label{fig:largedevices}
	\end{figure}

	In order to explore more in detail how the naive picture (the spin accumulation changing sign at the Dirac point) may be recovered, we have carried out TB calculations for systems up to $\approx 0.25\,\mu$m in length and for unrealistic large values of the Rashba coupling, equivalent to electric fields up to 2000 V/nm. Results for the armchair orientation are presented in Fig. \ref{fig:largedevices}. Top panels show the TB spin polarization including  only the Rashba term. In this case one can see the expected sign change in the spin polarization. The other panels show how the ``expected" result starts to appear at values of 20 V/nm when SOC is fully included. However, this requires a minimum length of the SOC-active region of $\approx 0.25\,\mu$m. As a summary, only for extremely large electric fields, the sign change can be observed for any length. This is a mesoscopic effect that can only be unveiled by resorting to calculations based on TB effective models.
	
	\section{Conclusions}
	
	We have studied the charge-spin interconversion capabilities of graphene from spin-orbit effects by means of DFT calculations. As expected, non-polarized spin currents injected in a region with Rashba SOC produce lateral spin accumulation at opposite edges (SHE) for the out-of-plane and longitudinal spin components and pure spin accumulation (REE) for the perpendicular spin component.
	
	To obtain the quantitative information we use a $k$-dependent spin polarization map obtained from DFT where the Rashba SOC was induced by an  electric field applied perpendicular to the graphene plane. A TB approach was additionally employed
	to verify our methodology, as an aid in the interpretation of the aforementioned charge-spin processes, and to overcome size limitations, inherent to DFT calculations, when deemed necessary. We have found that the longitudinal and perpendicular spin-projected current components are greater in absolute value than the out-of-plane component due to the nature of the Rashba effect.
	
	Moreover, we have found a competition between the intrinsic effect of the atomic SOC  and the Rashba effect (atomic SOC + broken inversion symmetry) which determines the sign of the pure spin accumulation for electrons. This can be changed by the length of the SOC active area or by the relative strength of the couplings. Our approach to these $k$-dependent studies can be applied to determine the different charge-spin interconversion processes in more complicated scenarios, such as when SOC effects are induced by proximity with other materials.
	
	\begin{acknowledgments}
		We acknowledge financial support through the “María de Maeztu” Programme for Units of Excellence in R\&D (CEX2018-000805-M), the Spanish MCIU and AEI and the European Union under Grants No. PID2019-109539GB-C43, PID2019-107874RB-I00 and PGC2018-097018-B-I00 (MCIU/AEI/FEDER, UE), the EU Graphene Flagship funding through JTC2017/2D-Sb\&Ge) and the Comunidad Autónoma de Madrid through Grants S2018/NMT-4321 (NanomagCOST-CM) and P2018/NMT-4411 (ADITIMAT-CM). J.J.P. and H.S. acknowledge the computer resources and assistance provided by the Centro de Computación Científica of the Universidad Autónoma de Madrid and the Red Española de Supercomputación. A.L. thanks the financial support of Brazilian CAPES/PrInt/UFF. 
	\end{acknowledgments}

\end{document}